\documentclass[journal]{IEEEtran}
\IEEEoverridecommandlockouts

\pdfoutput=1
\usepackage{cite}
\usepackage{amsmath,amssymb,amsfonts}
\usepackage{algorithmic}
\usepackage{caption}
\usepackage{textcomp}
\usepackage{balance}
\usepackage{subcaption}
\usepackage{algorithm,algorithmic}
\usepackage{hyperref}
\usepackage{changepage}

\usepackage[pdftex]{graphicx}
\usepackage{mdwmath}
\usepackage{eqparbox}
\usepackage{url}
\usepackage[inkscapeformat=png]{svg}

\def\BibTeX{{\rm B\kern-.05em{\sc i\kern-.025em b}\kern-.08em
    T\kern-.1667em\lower.7ex\hbox{E}\kern-.125emX}}
    
\begin{document}

\title{Alzheimer's disease detection in PSG signals}


\author{Lorena Gallego-Viñarás, Juan Miguel Mira-Tomás, Anna Michela Gaeta, Gerard Pinol-Ripoll, Ferrán Barbé, Pablo M. Olmos, Arrate Muñoz-Barrutia, \IEEEmembership{\textit{Fellow, IEEE}}
\thanks{Lorena Gallego-Viñarás, Juan Miguel Mira-Tomás and Arrate Muñoz-Barrutia are with the Bioengineering Department, Universidad Carlos III de Madrid, ES28911 Spain. Arrate Muñoz-Barrutia is also with the Instituto de Investigación Sanitaria Gregorio Marañón (IiSGM), Madrid, ES28007 Spain (e-mail: logalleg@pa.uc3m.es, 100383036@alumnos.uc3m.es, mamunozb@ing.uc3m.es).}
\thanks{Anna Michela Gaeta is with the Department of Pulmunology, Hospital Universitario Severo Ochoa, Leganés, ES28914 Spain (e-mail: annamichelagaeta@hotmail.it).}
\thanks{Gerard Pinol-Ripoll is with the IRBLleida-Hospital Universitari Santa Maria Lleida, Lleida, ES25198 Spain (e-mail: gpinol@gss.cat).}
\thanks{Ferrán Barbé is with the IRBLleida-Hospital Universitari Arnau de Vilanova and Santa Maria, Lleida, ES25198 Spain (e-mail: febarbe.lleida.ics@gencat.cat).}
\thanks{Pablo M. Olmos is with the Signal Processing Group (GTS), Universidad Carlos III de Madrid, Leganés, ES28911 Spain. Also, he is with the Instituto de Investigación Sanitaria Gregorio Marañón (IiSGM), Madrid, ES28007 Spain (e-mail: pamartin@ing.uc3m.es).}}

\maketitle
\begin{abstract}
Alzheimer's disease (AD) and sleep disorders exhibit a close association, where disruptions in sleep patterns often precede the onset of Mild Cognitive Impairment (MCI) and early-stage AD. This study delves into the potential of utilizing sleep-related electroencephalography (EEG) signals acquired through polysomnography (PSG) for the early detection of AD. Our primary focus is on exploring semi-supervised Deep Learning techniques for the classification of EEG signals due to the clinical scenario characterized by the limited data availability. The methodology entails testing and comparing the performance of semi-supervised SMATE and TapNet models, benchmarked against the supervised XCM model, and unsupervised Hidden Markov Models (HMMs). The study highlights the significance of spatial and temporal analysis capabilities, conducting independent analyses of each sleep stage. Results demonstrate the effectiveness of SMATE in leveraging limited labeled data, achieving stable metrics across all sleep stages, and reaching $90\%$ accuracy in its supervised form. Comparative analyses reveal SMATE's superior performance over TapNet and HMM, while XCM excels in supervised scenarios with an accuracy range of $92-94\%$. These findings underscore the potential of semi-supervised models in early AD detection, particularly in overcoming the challenges associated with the scarcity of labeled data. Ablation tests affirm the critical role of spatio-temporal feature extraction in semi-supervised predictive performance, and t-SNE visualizations validate the model's proficiency in distinguishing AD patterns. Overall, this research contributes to the advancement of AD detection through innovative Deep Learning approaches, highlighting the crucial role of semi-supervised learning in addressing data limitations.
\end{abstract}


\section{Introduction}
\label{sec:introduction}
\IEEEPARstart{T}{he} aging global population has led to an increase in chronic, age-related diseases. Alzeimer's disease  (AD), a neurodegenerative disorder affecting the central nervous system, is the most common form of dementia in older adults~\cite{WorldHealthOrganization},\cite{DONOSO2003}. AD is characterized by neuronal loss, brain atrophy, and the accumulation of beta-amyloid plaques and tau protein tangles\cite{proteinas_ad}.

Timely diagnosis of AD is crucial for implementing effective interventions. Current diagnostic practices primarily rely on observing clinical symptoms; however, emerging research indicates that sleep disturbances, especially Obstructive Sleep Apnea (OSA), may be early indicators of cognitive decline. OSA, known for causing breathing disruptions during sleep, is associated with an elevated risk of Mild Cognitive Impairment (MCI) and AD~\cite{ref9}. In light of this, polysomnography (PSG), a multifaceted sleep assessment method, has gained prominence in AD research~\cite{gaeta2020prevalence}. PSG captures a variety of physiological signals, including electroencephalography (EEG), offering valuable data on the sleep anomalies linked to AD~\cite{azami2023eeg}.

However, the inherent challenges in PSG studies, including cost and duration, contribute to the scarcity of comprehensive signal databases, particularly in AD patients.

\begin{figure*}[ht!]
    \centering
    \includegraphics[width=1\linewidth]{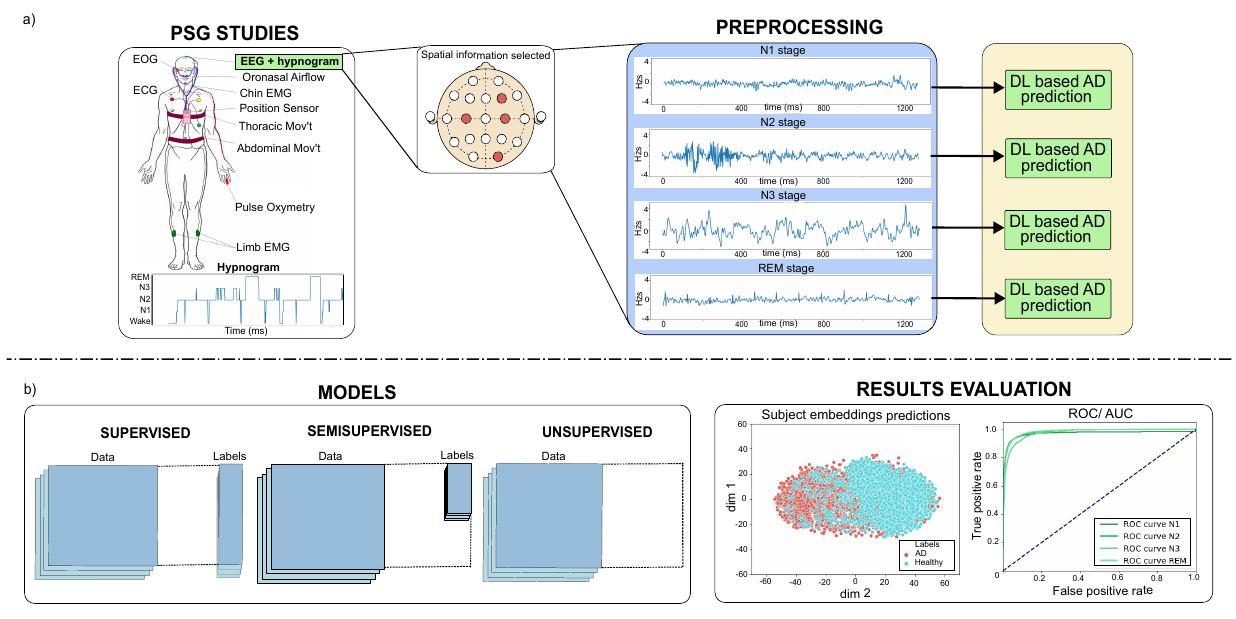} 
    \caption{(a) Block representation of the acquisition of PSG recordings and hypnograms, highlighting the spatial information extracted followed by the preprocessing block and the training and prediction of the different models in each sleep stage signals. (b) Types of models tested, representing the amount of data and the labels used, together with the block of predictions obtained for each supervised, semisupervised, and unsupervised model.}
    \label{graphical_abstract}
\end{figure*}
 
This manuscript builds upon previous research correlating sleep disorders with Alzheimer's disease and proposes an innovative framework for early AD detection using Deep Learning (DL) over sleep EEG signals. Recognizing the common clinical scenario characterized by constrained labeled data, our approach aims to propel the existing state-of-the-art forward by investigating semi-supervised DL models, as illustrated in Fig.~\ref{graphical_abstract}, mitigating the challenges of data scarcity and the laborious nature of conventional EEG analysis by:

\begin{itemize}
\item Testing and comparing the performance of semi-supervised DL models that effectively utilize both labeled and unlabeled EEG data against state-of-the-art supervised and unsupervised models.
\item Exploiting the temporal and spatial analysis capabilities of these models, crucial for understanding physiological signals and their complex dynamics~\cite{7488219}.
\item Conducting independent analyses of each sleep stage to pinpoint specific AD patterns~\cite{6845193}, acknowledging the distinct characteristics and frequencies of each phase. 
\end{itemize}


\begin{table*}[ht!]
    \centering
    \scalebox{1.2}{
        \begin{tabular}{| c | c | c | c | c |} \hline
            \textbf{Variables} & \textbf{AD Dataset} & \textbf{ISRUC Dataset} & \textbf{DOD-H Dataset} & \textbf{SDRC Dataset}\\ \hline
            \textbf{Age (years)} & 74.69 ± 5.06 & 40.00 ± 10.00 & 35.32 ±7.51 & 42.36 ± 15.74  \\
            \textbf{Gender (\% male)} & 50.81 & 90.00 &68.42 & 54.55 \\
            \textbf{Total Sleep Time (hours)} & 5.95 ± 2.93 & 7.16 ± 0.63 & 7.17 ± 1.11 & 6.70 ± 1.29 \\
            \textbf{Sleep efficiency (\%)} & 86.71 ± 39.36 & 79.79 ± 9.45 & 87.00 ± 8.10& 85.73 ± 15.64  \\ \hline 
        \end{tabular}
    }
    \caption{Demographic, clinical, and PSG variables of the four databases of the study.}
    \label{tab:databases}
\end{table*}



This research underscores the potential of PSG signal analysis in AD detection and the efficacy of diverse methods in deciphering complex signal patterns. By enhancing the accuracy and efficiency of sleep EEG signal classification, our research aims to significantly contribute to early AD detection, potentially improving treatment outcomes.

The manuscript is organized as follows: Section~\ref{sec: materials} presents the datasets used in this study. Section~\ref{sec: methods} describes the methodology, including the semi-supervised DL model and data preprocessing techniques. Section~\ref{sec: results} discusses the results, and Section~\ref{sec: discussion} delves into their implications. Finally, Section~\ref{sec: conclusion} concludes the paper, summarizing key insights and suggesting future research directions.


\subsection{Related Works}

Given the intricate relationship between EEG signals and neurodegenerative diseases such as AD, several studies have explored different DL methodologies for predicting such conditions. For example, Tanveer \textit{et al.}~\cite{9440810} implemented a variety of EEG data from awake subjects, achieving notable classification accuracies exceeding $97\%$ for AD and $83\%$ for MCI. In contrast, Klepl \textit{et al.}~\cite{10271565} adopted an alternative approach, leveraging Gated Graph Convolutional Networks for AD detection, with a particular emphasis on inter-regional brain connectivity, yielding commendable accuracy and underscoring the significance of spatial information from electrodes.

In a recent contribution, Tăuţan \textit{et al.}~\cite{TAUTAN2021102081} conducted a comprehensive review delineating the role of artificial intelligence in the detection of neurodegenerative diseases, specifically accentuating sleep signals as potential biomarkers for various of these neurodegenerative conditions. Moreover, in 2021, D'Atri \textit{et al.}~\cite{DAtri_Scarpelli_Gorgoni_Truglia_Lauri_Cordone_Ferrara_Marra_Rossini_De_Gennaro_2021} established the presence of statistically significant differences among various frequency bands and several electrodes in both sleep and wake signals across healthy individuals, those with MCI, and AD patients.

In line with this perspective, the analysis of sleep signals has emerged as a significant area of study. While some efforts have been made in AD detection during these sleep signals, such as Geng \textit{et al.}'s  \cite{Geng_Wang_Fu_Zhang_Yang_An_2022} implementation of GRUs based on features extracted from sleep EEG, and Azami \textit{et al.}'s \cite{Azami_Moguilner_Penagos_Sarkis_Arnold_Gomperts_Lam_2023} study on entropy measurements in specific sleep stages, there remains a significant research gap in the thorough exploration of PSG-derived signals for AD patients. This gap is largely due to the inherent complexity of these tests and the limited availability of data. This current study seeks to address this gap, focusing on advancing the analysis and utilization of sleep EEG signals for the early detection of AD.

\section{Materials}
\label{sec: materials}

This study leverages four fully labeled databases chosen for their diversity and relevance to Alzheimer's disease (AD) and sleep study. 

\subsection{Alzheimer Disease Patient Database}

The AD Database includes data from $58$ individuals diagnosed with mild to moderate AD, aged $60$ and above. This data, collected under the clinical trial NCT02814045 at Santa Maria University Hospital in Lleida, Spain, between $2015$ and $2017$, comprises overnight PSG recordings. The recordings include six EEG signals from both hemispheres, but for our analysis, we focus on four channels (EEGC3-A2, EEGC4-A1, EEGF4-A1, and EEGO2-A1), as recommended by the AASM~\cite{ASSM}. This selection aligns with findings on reduced inter-hemispheric coherence in AD patients~\cite{27}. Artifacts identified in these signals have been marked for exclusion. Additionally, each recording includes an annotated hypnogram detailing sleep phases.

\subsection{Healthy Controls Database}

To compare AD patient data with healthy individuals without any cardiac, neurological, or psychiatric disorders, we compiled a database from three public sources: 

\begin{itemize}
    \item \textbf{ISRUC-SLEEP Dataset (Subgroup-III):} Comprising data from 10 subjects (9 males, 1 female) aged around 40 years. Four EEG channels (C3-A2, C4-A1, F4-A1, O2-A1) were selected, filtered (0.3-35 Hz), and sampled at 200 Hz~\cite{30}.

    \item \textbf{Dream Open Dataset-Healthy (DOD-H)}: Encompassing 25 individuals, this dataset excludes subjects with recent severe health conditions. EEG data from six channels recorded on a Siesta 802 device were sampled at 256 Hz and filtered (0.03-35 Hz). For our study, four channels (C3/M2, C4/M1, F4/M1, and O2/M1) were used~\cite{dodh}.

    \item \textbf{Sleep Disorders Research Center (SDRC) Dataset}: Featuring 11 subjects from Kermanshah, Iran, without sleep disorders. Data was collected using a SOMNOscreen™ plus PSG device at a 256 Hz sampling rate. From the 14 recorded EEG channels, we selected C4A1, C3A2, F4A1, and O2A1~\cite{32}.
    
\end{itemize}

Table \ref{tab:databases} details the demographic and clinical characteristics of these combined databases. It is important to note that all EEG signals are monopolar, ensuring consistency across databases. Each dataset also includes expert-extracted hypnograms. Furthermore, it has been shown that the parietal region of the brain could manifest the most noticeable age-dependent differences~\cite{Terry_Anderson_Horne_2004}. This justifies the choice of electrodes from different brain areas to minimize the bias in our study in response to the age gap between the merged databases. In addition, our experiments show that the patterns associated with AD in the signals are more pronounced than those linked to age.

\begin{figure*}[ht!]
    \centering
    \includegraphics[width=0.9\linewidth]{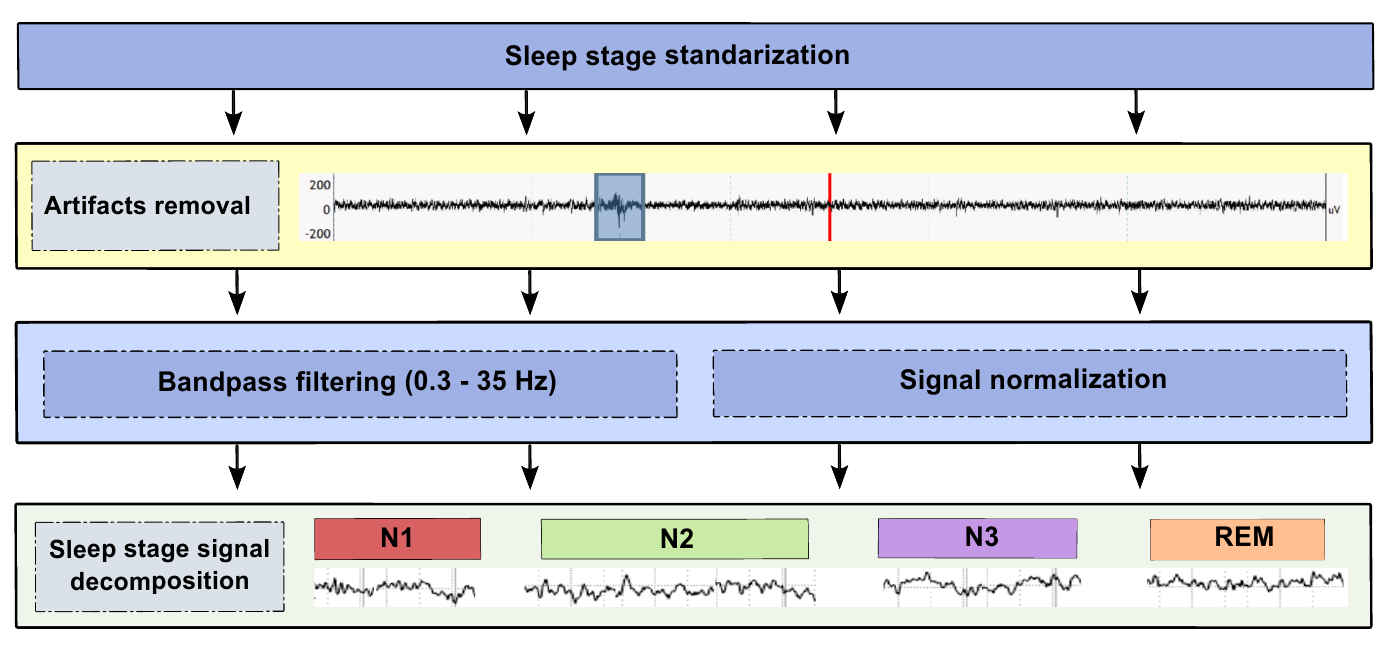} 
    \caption{Workflow of EEG Signal Preprocessing. This figure illustrates the step-by-step transformation of raw EEG data from various databases into standardized, harmonized segments. It outlines the processes of hypnogram standardization, artifact removal, signal filtering, normalization, segmentation by sleep stages, and final resampling and segmentation into uniform 10-second segments. This flowchart demonstrates how disparate data sources are methodically processed to ensure consistency and reliability for subsequent analysis.}
    \label{plan1}
\end{figure*}

\section{Methods}
\label{sec: methods}

\subsection{Preprocessing of EEG Signals}

Given the varied characteristics of the EEG signals sourced from four distinct databases, harmonizing these data was a critical initial step in our study. This preprocessing ensures consistency, enabling reliable analysis and comparison. 

\textbf{Standardization of Hynograms:} We began by standardizing the hypnograms to align with the current sleep stage nomenclature. Older scoring rules classified sleep into seven stages, but we adopted the current system recognizing three Non-REM (N1, N2, N3) phases, REM and Wake stages. The N3 stage comprises the former S3 and S4 stages. Additionally, what was previously categorized as 'movement time' is now recognized as artifacts~\cite{S-to-N}.

\textbf{Artifact Annotation and Removal}: Variations in artifact annotations across databases necessitated careful review and removal to avoid any distortions in EEG signal analysis.

\textbf{Signal Filtering}: To harmonize signal quality, we applied a Butterworth bandpass filter (0.3-35 Hz) to the AD and SDRC databases. This filter was chosen for its smooth passband response and minimal phase distortion, which is important for maintaining the integrity of EEG signals. The filter's range was set to preserve essential EEG frequencies, from the Delta band (around 0.5 Hz) to the Beta band (up to 32 Hz)~\cite{waves_freq}. EEG signals from other databases had already been filtered within this range.

\textbf{Normalization and Segmentation}: We normalized all signals to a zero mean and a standard deviation of one, reducing variability across databases and enhancing numerical stability. Following this, signals were segmented according to the corresponding sleep stages from the hypnograms. This segmentation led to the creation of four separate databases, each representing one of the sleep stages: N1, N2, N3, and REM. In our study, sleep phase annotations were manually conducted by clinicians, notwithstanding the existence of various automated methods available for performing this task.

\begin{table}[t!]
    \centering
    \scalebox{1.25}{
        \begin{tabular}{| c | c | c | c | c | } \hline
            \textbf{Sleep Stage} & \textbf{N1}  & \textbf{N2} & \textbf{N3} & \textbf{REM}\\ \hline
            \textbf{HC dataset} & 18459 & 49287 & 20685 & 19485 \\
            \textbf{AD dataset} & 17120 & 31944 & 20011 & 9771  \\ \hline
        \end{tabular}
    }
    \caption{Final number of signal fragments present for each sleep stage after preprocessing.}
    \label{tab:sleepstages_division}
\end{table}

\begin{figure*}[t!] 
    \centering
    
    \begin{subfigure}{0.49\textwidth}
        \centering
        \includegraphics[width=\textwidth]{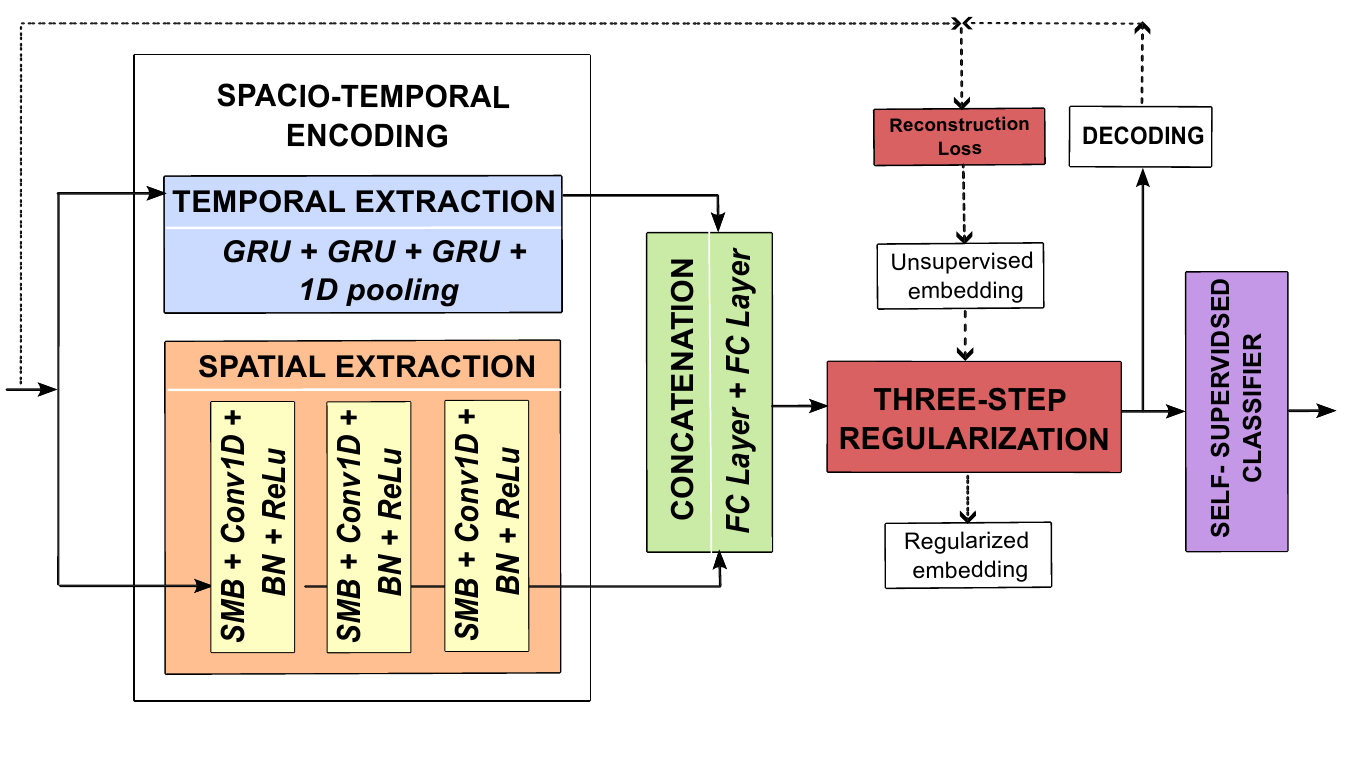}
        \caption{}
        \label{smate_fig}
    \end{subfigure}
    \hfill 
    \begin{subfigure}{0.49\textwidth}
        \centering
        \includegraphics[width=\textwidth]{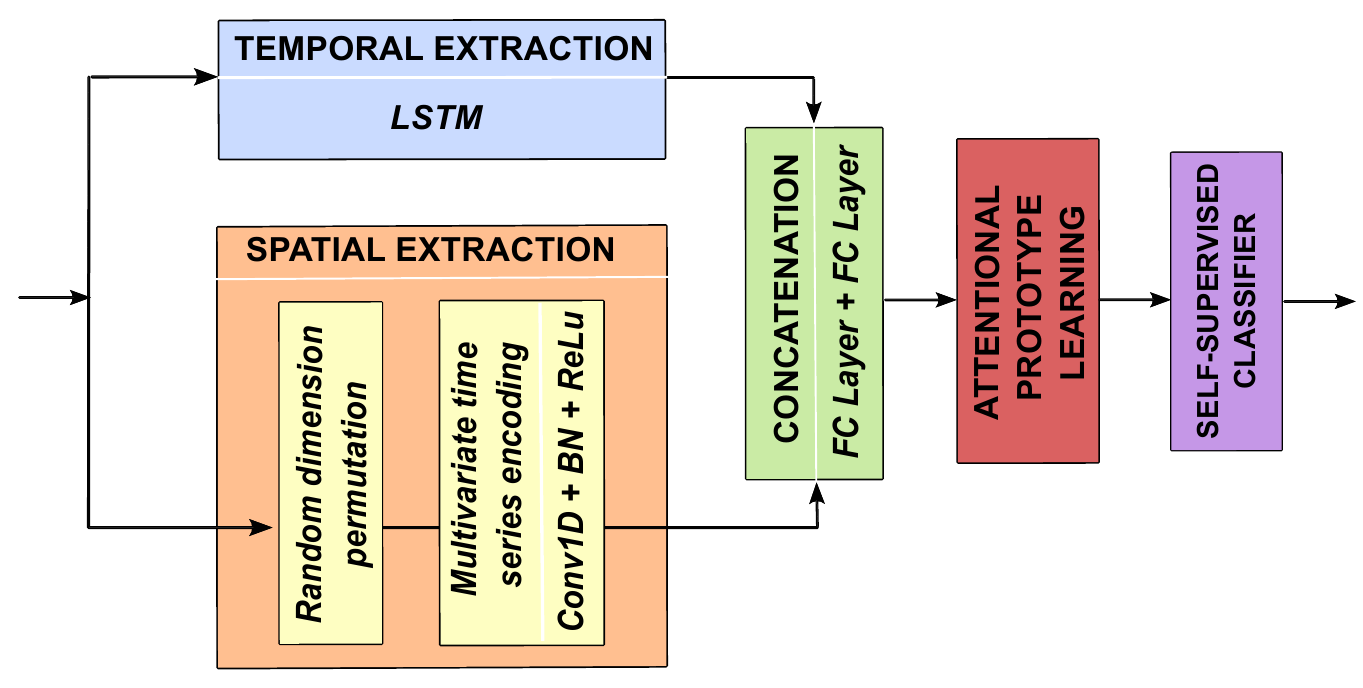}
        \caption{}
        \label{tapnet_fig}
    \end{subfigure}
    
    \medskip 
    
    \begin{subfigure}{0.49\textwidth}
        \centering
        \includegraphics[width=\textwidth]{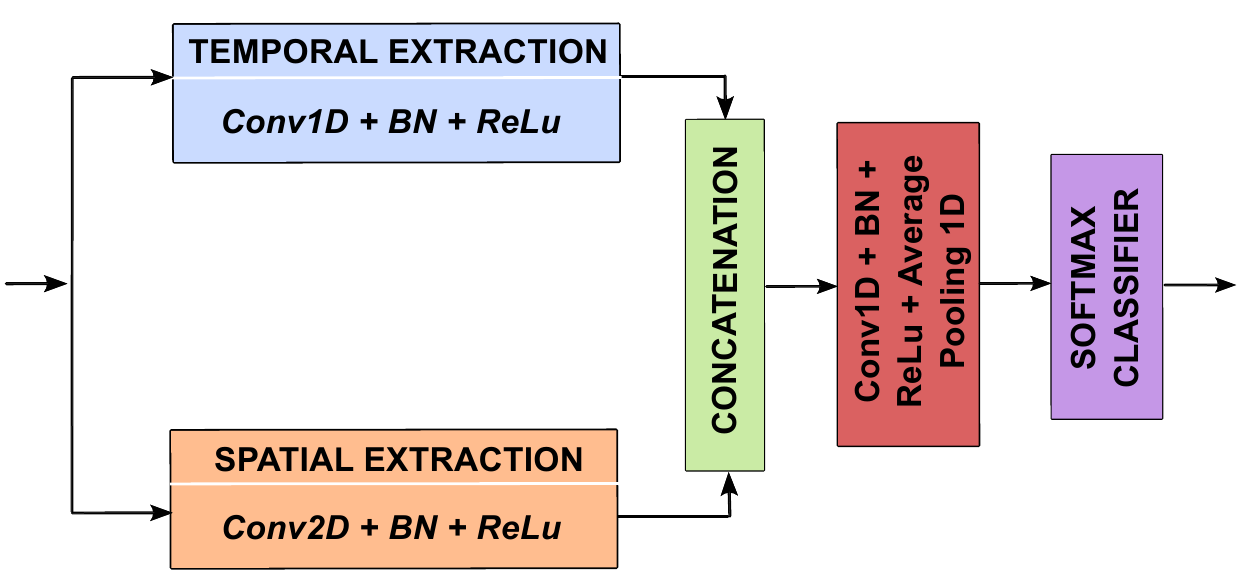}
        \caption{}
        \label{xcm_fig}
    \end{subfigure}
    \hfill 
    \begin{subfigure}{0.49\textwidth}
        \centering
        \includegraphics[width=\textwidth]{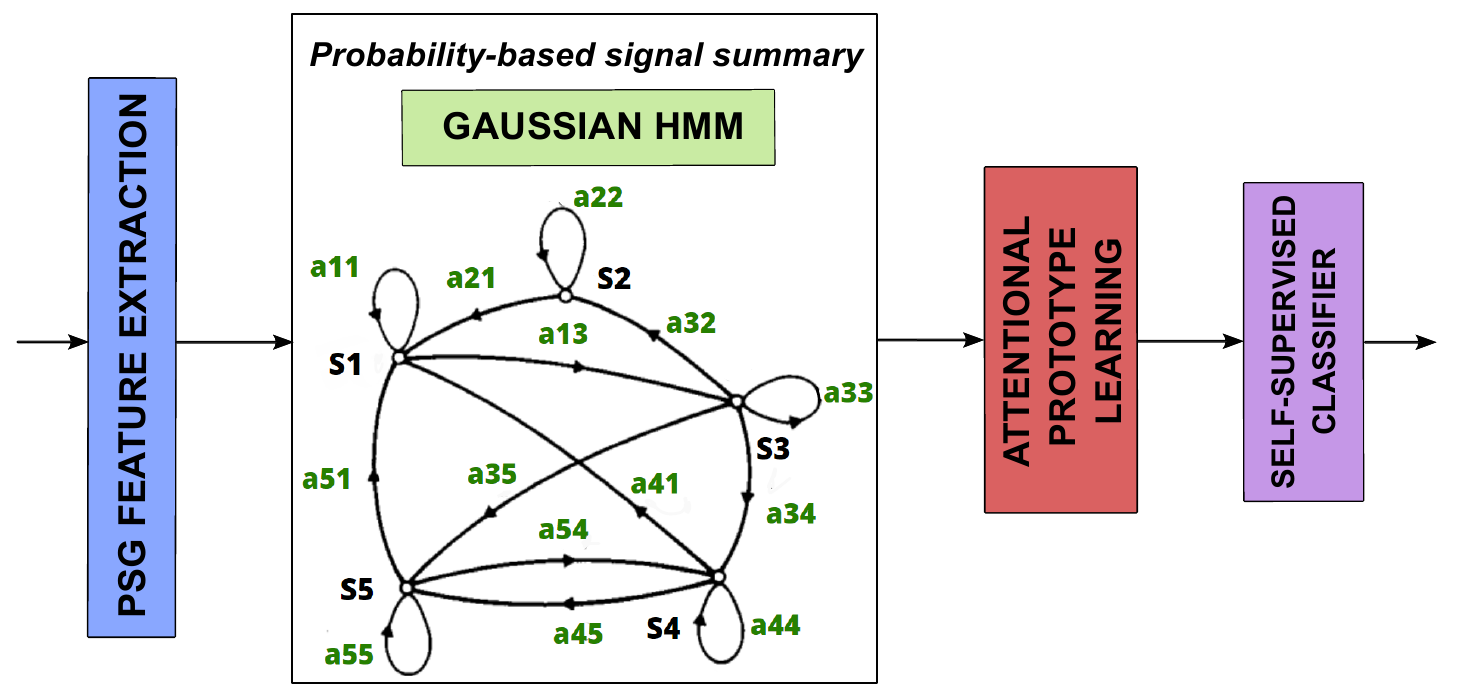}
        \caption{}
        \label{hmm_fig}
    \end{subfigure}
    
    \caption{Illustrative Overview of Model Structures Employed. This figure provides a schematic representation of each model's architecture. (a) The SMATE model, focusing on its integration of spatiotemporal features, detailed regularization processes, and the decoding block for final classification~\cite{smate}; (b) The TapNet model, showcasing its sequential information extraction process, the innovative random dimension permutation, the encoding of multivariate time series, and its unique attentional prototype learning mechanism~\cite{tapnet}; (c) The XCM model, emphasizing its dual approach with 1D temporal feature extraction and 2D spatial feature extraction, offering a holistic view of the data~\cite{xcm}; (d) The structure of the HMM model, describing its approach to PSG feature extraction, and detailing the implementation of a 5-state Hidden Markov Model to keep a balance between performance and computational cost, including state transitions for sequence analysis.}
    \label{all_models}
\end{figure*}

\begin{figure}[ht!]
    \centering
    \includegraphics[width=1\linewidth]{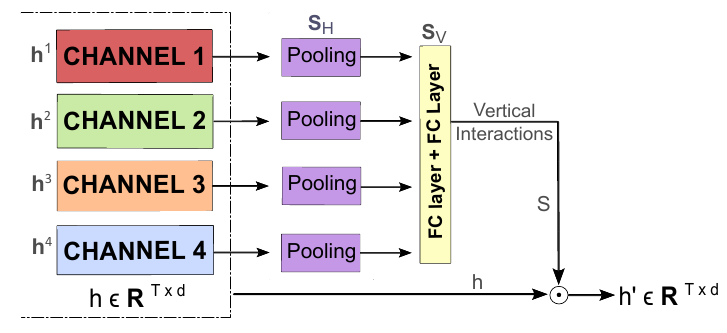} 
    \caption{Spatial Modeling Block (SMB) architecture (adapted from \cite{smate}).}
    \label{smb}
\end{figure}

\textbf{Signal Resampling and Segmentation}: The resultant signals were resampled at 128 Hz, a decision aimed at reducing data complexity and computational demands. Further, we divided these signals into uniform 10-second segments, ensuring consistent lengths for comparative analysis across all databases. Table \ref{tab:sleepstages_division} provides detailed information on this division.

Following the completion of all pre-processing steps, the distinct signal fragments were concatenated based on their association with each patient's ID, facilitating subsequent division into training, validation, and test sets.
Figure~\ref{plan1} presents a visual summary of these preprocessing steps, illustrating the transformation from raw data to standardized segments ready for analysis.

\subsection{EEG classification with DL}

This study employs a range of deep learning models to analyze multivariate time series (MTS) EEG data, each with unique approaches to feature extraction and classification. Figure~\ref{all_models} provides an overview of their architectures. Building upon the findings of~\cite{DAtri_Scarpelli_Gorgoni_Truglia_Lauri_Cordone_Ferrara_Marra_Rossini_De_Gennaro_2021}, our methods seek to extract both spatial and temporal characteristics to acquire more comprehensive information from the signals. Additionally, various proportions of labeled and unlabeled data were examined to evaluate the performance of the semi-supervised models.

\subsubsection{\textbf{SMATE Model}}

SMATE (Semi-Supervised Spatio-Temporal Representation Learning on MTS)~\cite{smate} is designed for weakly labeled Multivariate Time Series (MTS) analysis. It uses an autoencoder framework to compress MTS data into a condensed embedding space. This process allows SMATE to capture spatial and temporal dynamics using various blocks, with a focus on integrating both labeled and unlabeled data for model optimization. The key components of SMATE include a Spatial Modeling Block (SMB) for EEG channel interactivity analysis, a temporal extraction block incorporating Gate Recurrent Units (GRUs), spatiotemporal encoding, and a joint model optimization strategy. This strategy involves a three-step regularization process to adress the sparse distribution of the autoencoder framework, refining class-specific clusters for each embedding space and integrating labeled and unlabeled data for enhanced learning (see Fig.~\ref{smate_fig} and \ref{smb}). 

The three-step regularization process comprises a supervised centroid initialization, defining each class centroid with its specific embeddings. This is followed by a supervised centroid adjustment, refining centroids using probabilities derived from sample distances to each class. The final stage is unsupervised centroid adjustment, where unlabeled samples contribute to fine-tuning the centroids. The regularization loss is computed based on weights associated with labeled samples or distance-based probabilities of unlabeled samples. 

\subsubsection{\textbf{TapNet Model}}

TapNet~\cite{tapnet}, or 'Multivariate Time Series Classification with Attentional Prototypical Network,' is a semi-supervised model that utilizes an attentional network for feature representation. This network focuses on learning from proximity to class prototypes, making it effective in scenarios with sparse training labels. The architecture includes Long Short-Term Memory (LSTM) networks for extracting contextual embeddings, a random dimension permutation component, and a robust convolutional network for encoding time series. The model's attentional prototype learning phase leverages unlabeled data, enhancing training and estimation in label-scarce environments (refer to Fig.~\ref{tapnet_fig})

\subsubsection{\textbf{XCM Model}}

XCM~\cite{xcm} is a supervised model that extracts parallel information from observed variables and time dimensions. It utilizes 1D and 2D convolutional filters for feature extraction and simplifies the classification process by employing 1D average pooling before prediction. This approach reduces parameter count and improves generalization, eliminating the need for Fully Connected (FC) layers (see Fig.~\ref{xcm_fig}).

\subsection{Hidden Markov Models}

Hidden Markov Models (HMMs) are used for their proficiency in modeling time-dependent data. They are unsupervised statistical models that analyze sequences where observable events are linked to underlying hidden states~\cite{git}. In our study, HMMs analyze EEG signals post-feature extraction to detect patterns and classify data. This process involves an attentional prototype learning as final step to perform the final classification (Fig.~\ref{hmm_fig}).

\subsection{Statistical analysis}

An Analysis of Variance (ANOVA) test~\cite{anova} was employed to analyze the results statistically. This test compares the variance between group means against the average variance within each group. To evaluate the results, we performed comparative analyses among the different models: SMATE (both unsupervised and supervised), TapNet (unsupervised and supervised), XCM (supervised), and HMM (unsupervised). Additionally, we compared the performance of SMATE and TapNet across varying levels of labeled data, ranging from 0\% to 100\% in increments of 10\%. These comparisons were also extended to different sleep stages, specifically N1, N2, N3, and REM. For all these analyses, we set the confidence interval at 0.05. 


\section{Results}
\label{sec: results}
The analyses were performed using an NVIDIA GeForce RTX 2090 GPU, equipped with 24 GB of memory and running CUDA 11.6, to ensure efficient data processing.

In assessing our objective of early Alzheimer's disease detection, we have examined the performance of semi-supervised models. Furthermore, we conducted a comparative analysis between the semi-supervised SMATE and TapNet, alongside the supervised XCM and unsupervised HMM. This comparison aims to highlight not just numerical outcomes but also differences in robustness and stability, as well as the potential advantages and drawbacks of semi-supervised models.

\begin{table*}
    \centering
    
    \scalebox{1.2}{ 
        \begin{tabular}{| c | c | c | c | c |} \hline
            \textbf{Model} & N1 & N2 & N3 & REM  \\ \hline
            \multicolumn{5}{|c|}{\textbf{supervised}}\\ \hline
            \textbf{SMATE} & 0.919 ± 0.015 & 0.905 ± 0.030 & 0.871 ± 0.029 & 0.921 ± 0.007  \\
            \textbf{TapNet} & 0.784 ± 0.081* & 0.663 ± 0.056* & 0.684 ± 0.085* & 0.797 ± 0.053* \\
            \textbf{XCM} & \textbf{0.94 ± 0.023} & \textbf{0.946 ± 0.024} & \textbf{0.943 ± 0.023} & \textbf{0.941 ± 0.024} \\ \hline

            \multicolumn{5}{|c|}{\textbf{unsupervised}}\\
            \hline 

            \textbf{SMATE} & \textbf{0.739 ± 0.087} & 0.66 ± 0.076 & \textbf{0.67 ± 0.09} & 0.6838 ± 0.094\\
            \textbf{TapNet} & 0.708 ± 0.105 & 0.656 ± 0.075 & 0.606 ± 0.090 & \textbf{0.697 ± 0.103} \\
            \textbf{HMM} & 0.684 ± 0.094 & \textbf{0.699 ± 0.078} & 0.642 ± 0.098  & 0.635 ± 0.102  \\ \hline 
            
            \multicolumn{5}{|c|}{\textbf{supervised no spatial block (accuracy)}}\\ \hline
            \textbf{SMATE} & 0.895 ± 0.017* & 0.887 ± 0.035* & 0.827 ± 0.017* & 0.871 ± 0.04*  \\
            \textbf{TapNet} & 0.538 ± 0.07* & 0.475 ± 0.071* & 0.509 ± 0.045* & 0.469 ± 0.057* \\
            \textbf{XCM} & \textbf{0.932 ± 0.023*} & \textbf{0.96 ± 0.019*} & \textbf{0.949 ± 0.019*} & \textbf{0.916 ± 0.049*} \\ \hline
            
            \multicolumn{5}{|c|}{\textbf{supervised no spatial block (ROC/AUC)}}\\
            \hline 

            \textbf{SMATE} & 0.937 ± 0.013* & 0.854 ± 0.037* & 0.845 ± 0.04* & 0.789 ± 0.042*  \\
            \textbf{TapNet} & 0.50 ± 0.0* & 0.50 ± 0.0* & 0.50 ± 0.0* & 0.50 ± 0.0* \\
            \textbf{XCM} & \textbf{0.990 ± 0.005*} & \textbf{0.994 ± 0.004*} & \textbf{0.989 ± 0.005*} & \textbf{0.986 ± 0.004*} \\ \hline
        
            \multicolumn{5}{|c|}{\textbf{supervised no temporal block (accuracy)}}\\ \hline
            \textbf{SMATE} & \textbf{0.901 ± 0.023} & \textbf{0.869 ± 0.021} & 0.85 ± 0.013 & 0.874 ± 0.018  \\
            \textbf{TapNet} & 0.791 ± 0.064 & 0.631 ± 0.080 & 0.634 ± 0.102 & 0.756 ± 0.101 \\
            \textbf{XCM} & 0.708 ± 0.272 & 0.808 ± 0.186 & \textbf{0.862 ± 0.186} & \textbf{0.916 ± 0.049} \\ \hline

            \multicolumn{5}{|c|}{\textbf{supervised no temporal block (ROC/AUC)}}\\
            \hline 
            \textbf{SMATE} & \textbf{0.937 ± 0.013} & \textbf{0.952 ± 0.028} & 0.918 ± 0.009 & \textbf{0.937 ± 0.015}  \\
            \textbf{TapNet} & 0.691 ± 0.234* & 0.50 ± 0.0* & 0.50 ± 0.0* & 0.50 ± 0.0* \\
            \textbf{XCM} & 0.888 ± 0.194 & 0.892 ± 0.195 & \textbf{0.976 ± 0.015} & 0.787 ± 0.234 \\ \hline
        \end{tabular}}
            \caption{Performance Comparison of Models on PSG Dataset: This table presents a comparative analysis of test accuracy and standard deviation (averaged over 5 folds) for each model applied to the PSG dataset. The test results are segmented by sleep stages (N1, N2, N3, and REM). An asterisk (\textbf{*}) next to the accuracy values denotes stages where the performance shows statistically significant differences (with a $p$-value $<$ 0.05) compared to the corresponding stages of other models. This distinction provides insights into the relative effectiveness of each model across different sleep stages and through the different ablation tests.}
    \label{tab:test_results}
\end{table*}

Utilizing a 5-fold cross-validation ensured broad generalization and a stringent evaluation. The dataset division was 70\% of the patients for training, 10\% for validation, and 20\% for testing. For SMATE and TapNet, tests were conducted with varying proportions of labeled samples, from 0\% to 100\%. To ensure comparability, tests with 100\% labeled samples were also conducted for all models. For SMATE, a minimal percentage of labeled samples was used, as its method requires a non-zero number for centroid initialization.

For this study, all models were trained using the Adam optimizer, set at a learning rate of $10^{-5}$, and a batch size of 128. Specific parameters for each model are detailed in Appendix~\ref{FirstAppendix}.


\subsection{Comparison between fully supervised vs. unsupervised models}

Table~\ref{tab:test_results} shows the results, using Accuracy as the classification metric, between supervised (SMATE, TapNet and XCM) and unsupervised methods (SMATE, TapNet and HMM). The results are separated by sleep phases.

\begin{figure*}[htbp]
    \centering
    
    \begin{subfigure}{1\linewidth}
        \includegraphics[width=\linewidth]{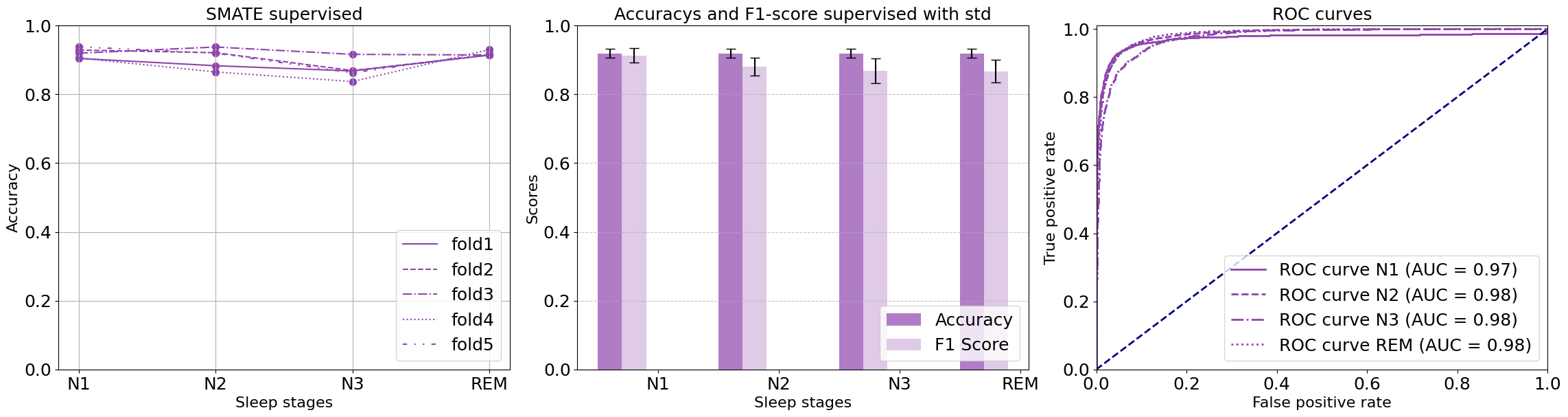}
        \caption{}
        \label{subfig:smate_res}
    \end{subfigure}
    \hfill
    \begin{subfigure}{\linewidth}
        \includegraphics[width=\linewidth]{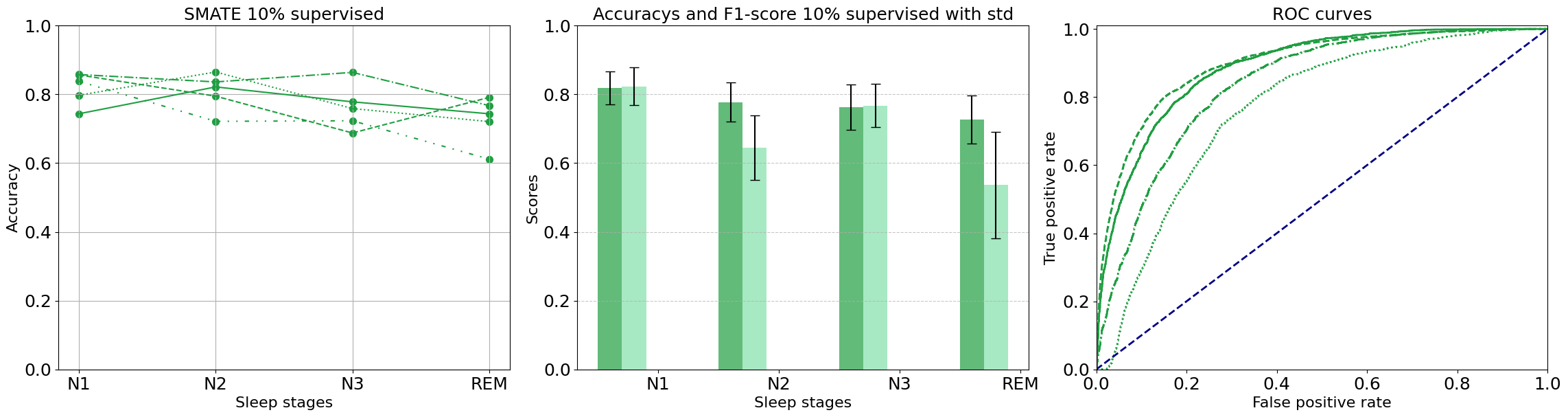}
        \caption{}
        \label{subfig:smate_01_res}
    \end{subfigure}    
    \begin{subfigure}{1\linewidth}
        \includegraphics[width=\linewidth]{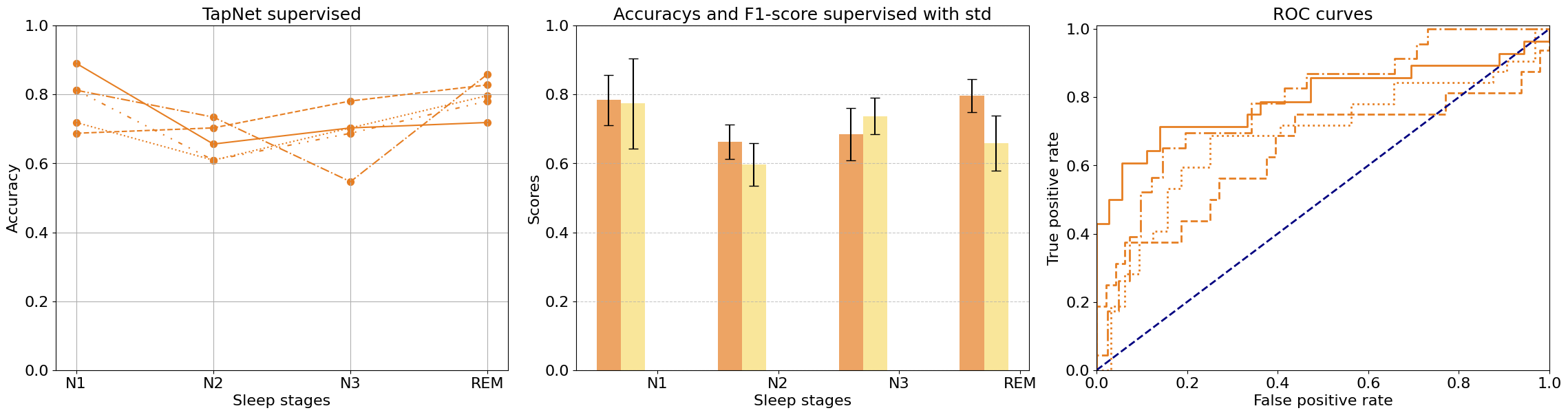}
        \caption{}
        \label{subfig:tapnet_res}
    \end{subfigure}
    \caption{Comparative analysis of SMATE supervised, SMATE with 10\% labeled samples and TapNet supervised model performances. This figure displays the accuracy, F1-score metrics (with their standard deviations), and ROC curves for each model across different sleep stages. Each panel, from left to right, represents the results of each fold for the corresponding model: (a) Depicts the SMATE fully supervised model, showcasing its performance consistency across sleep stages; (b) Illustratres the SMATE 10\% supervised model,  showcasing its variability across different stages while still outperforming (c), which represents the TapNet fully supervised model, revealing greater variability compared to SMATE across the sleep stages.} 
    \label{combined_results}  
\end{figure*}

The XCM model exhibited superior performance, achieving 92-94\% accuracy across sleep stages, with minimal standard deviation (Table~\ref{tab:test_results}). SMATE demonstrated elevated and consistent metric values through all sleep stages, yet it did not surpass XCM in its supervised mode (Fig.\ref{subfig:smate_res}). In its unsupervised mode, SMATE maintains a similar performance compared to both TapNet and HMM. TapNet's accuracy hovered around 70\%, showing greater variability and a marked drop in performance during the N3 phase (Fig.\ref{subfig:tapnet_res}). The HMM model lagged behind, averaging around 65\% accuracy (Table~\ref{tab:test_results}). For HMM, a Gaussian model with five hidden states and four emissions was chosen, optimizing features such as mean, standard deviation, kurtosis, and Hjorth parameters.

ROC curves were plotted for each model and sleep stage, indicating the trade-off between True Positive Rate and False Positive Rate at different thresholds (Fig.~\ref{combined_results}). In its supervised configuration, SMATE displayed near-perfect curves. However, when trained with only 10\% of the samples labeled, there was a significant decrease in performance. Despite this decline, it still outperformed TapNet in its supervised mode, which exhibited inconsistent curves, notably in the N3 stage. This observation suggests the robustness of the SMATE model.

To assess the potential bias arising from varying age ranges among patients, tests were conducted to evaluate its impact. Three categories were defined, encompassing patients aged 30-47 years in the first group, 48-69 in the second, and 70-84 in the third, each group consisting of an equal number of individuals. Following this categorization, the models were re-trained to classify signals within each new age category. The tests were unsuccessful, as none of the metrics surpassed 0.4 in any sleep stage.

\textbf{Statistical Analysis:} An ANOVA test revealed statistically significant differences among the models. The post-hoc analysis identified these discrepancies, particularly between SMATE and XCM, which have higher performance than TapNet. In the unsupervised comparison, SMATE, TapNet and HMM showed no statistically significant differences, exhibiting similar performance in all sleep phases.


\subsection{Ablation test}
The elimination of the spatial block yielded significant changes in the models' capability for processing contextual information as depicted in Table~\ref{tab:test_results}. In the semi-supervised models, the absence of spatial information resulted in a noticeable decrease in both consistency and coherence. Specifically, TapNet was the most affected model, exhibiting a random classification considering the ROC/AUC metrics. Similarly, SMATE demonstrated reduced metric values across all phases, yet the decline was less severe compared to TapNet, with REM sleep showing the most significant decrease. Meanwhile, the supervised XCM model showed considerable robustness against the removal of spatial information, with only a minor impact on its overall performance.  

Moreover, eliminating the temporal block also led to changes in the classification abilities of the models (Table~\ref{tab:test_results}). In this case, the semi-supervised SMATE model experiences a drop in performance, albeit less pronounced than observed in the spatial block ablation scenario. As occurs in the combination of the spatial and temporal block, N2 sleep phase is the most affected. TapNet also exhibited stochastic classification in the N2, N3, and REM phases when the temporal block was removed.

Conversely, the omission of the temporal block significantly diminished the analytical capabilities of the supervised XCM model across all stages, notably failing to reach a ROC/AUC metric value of 0.8 in the REM phase. 

\textbf{Statistical Analysis:} ANOVA test revealed statistically significant differences between models, especially in the spatial block ablation. Post-hoc analysis identified these discrepancies between all the SMATE, TapNet and XCM models, both in terms of Accuracy and ROC/AUC, across all sleep phases. In the case of temporal block ablation, TapNet differed significantly from SMATE and XCM, revealing an inferior performance to these models in every sleep stage, notably in terms of ROC/AUC.

\subsection{Semi-supervised models performance}

\begin{figure}[htbp]
    \begin{adjustwidth}{-0.in}{0in}
        \centering
        \begin{subfigure}[b]{1\linewidth} 
            \includegraphics[width=\linewidth]{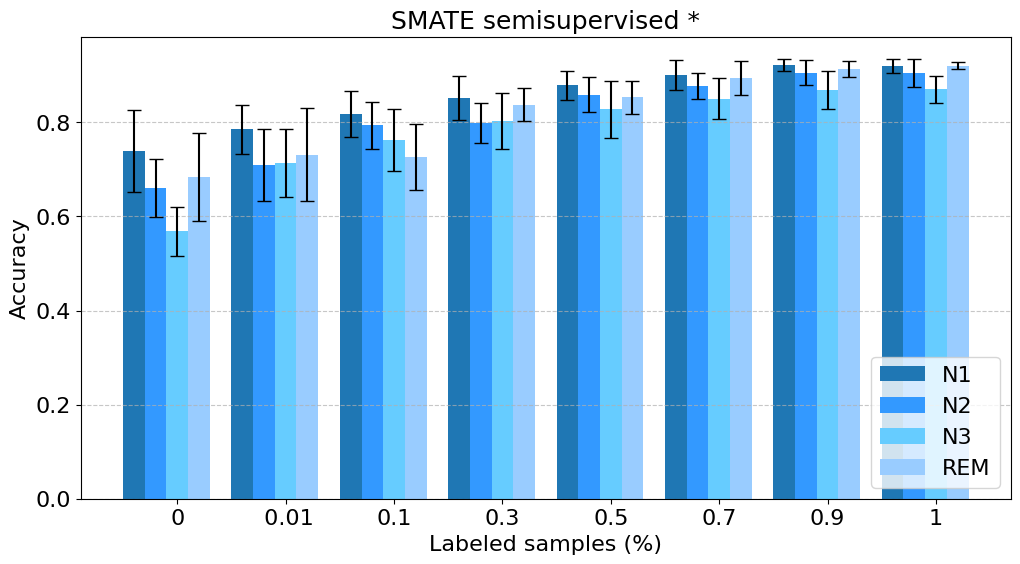}
            \caption{}
            \label{semisup_smate}
        \end{subfigure}
        
        \begin{subfigure}[b]{1\linewidth} 
            \includegraphics[width=\linewidth]{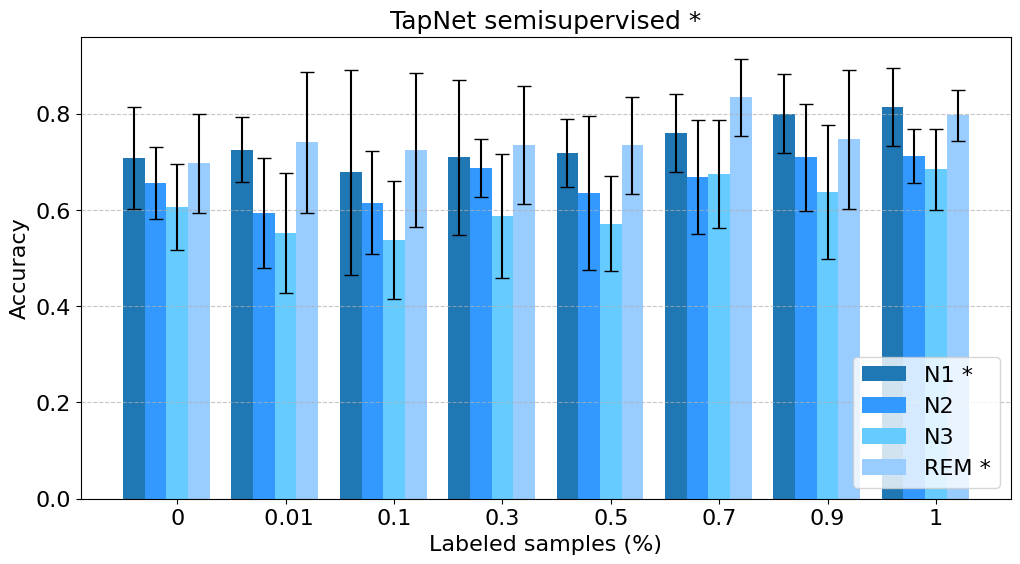}
            \caption{}
            \label{semisup_tapnet}
        \end{subfigure}
        
        \caption{Accuracy (mean and std) comparison of the semi-supervised SMATE (a) and TapNet (b) models, varying the percentage of labeled samples from 0\% to 100\%. The asterisk (*) in SMATE and TapNet models indicates statistically significant differences within all stages $(p < 0.05)$ with this percentage variation.}
        \label{fig:semisupervised}
    \end{adjustwidth}
\end{figure}

In evaluating the semi-supervised models SMATE and TapNet, both were tested with various proportions of labeled data (from 0\% to 100\% in 20\% intervals). While TapNet and SMATE exhibit comparable performance in their unsupervised states, SMATE demonstrates notable and consistent improvements in accuracy as the quantity of labeled samples increases. This stability across all sleep phases, along with significant accuracy enhancements with minor variations in labeled data, underscores SMATE's adeptness at extracting pertinent patterns from unlabeled data with this minimal additional information, as depicted in Fig.~\ref{semisup_smate}. Analysis of Variance (ANOVA) tests indicated no statistically significant disparities in performance across sleep stages, albeit visually, the N1 stage consistently outperforms others.  Notably, the training duration for SMATE decreased as the proportion of labeled data was reduced.

In contrast, the TapNet model showed a lower increasing trend in performance as supervision increased, from 0\% to 100\%, as depicted in Fig.~\ref{semisup_tapnet}. This trend suggests a decline in accuracy with reduced labeled data, particularly in the N3 phase, which consistently underperformed compared to the N1 and REM stages. This pattern indicates that TapNet is less resilient than SMATE when dealing with lesser amounts of labeled data. The ANOVA test further supports this, highlighting statistically significant differences in N1 and REM stages with respect to N2 and N3.

Both SMATE and TapNet showed statistically significant variations within each stage as the percentage of labeled samples changed.

\subsection{Visual Interpretation of the Embedding Space}

We employed t-distributed Stochastic Neighbor Embedding (t-SNE), an unsupervised, non-linear dimensionality reduction technique, to reveal the intricate structures in our high-dimensional DL models. T-SNE facilitates the visualization of the class distribution in our binary classification problem on a 2D plane, providing insight into how the models differentiate between classes~\cite{tsne2}.

\begin{figure*}[t!] 
    \begin{subfigure}{\textwidth}
        \centering
        \includegraphics[width=0.77\textwidth]{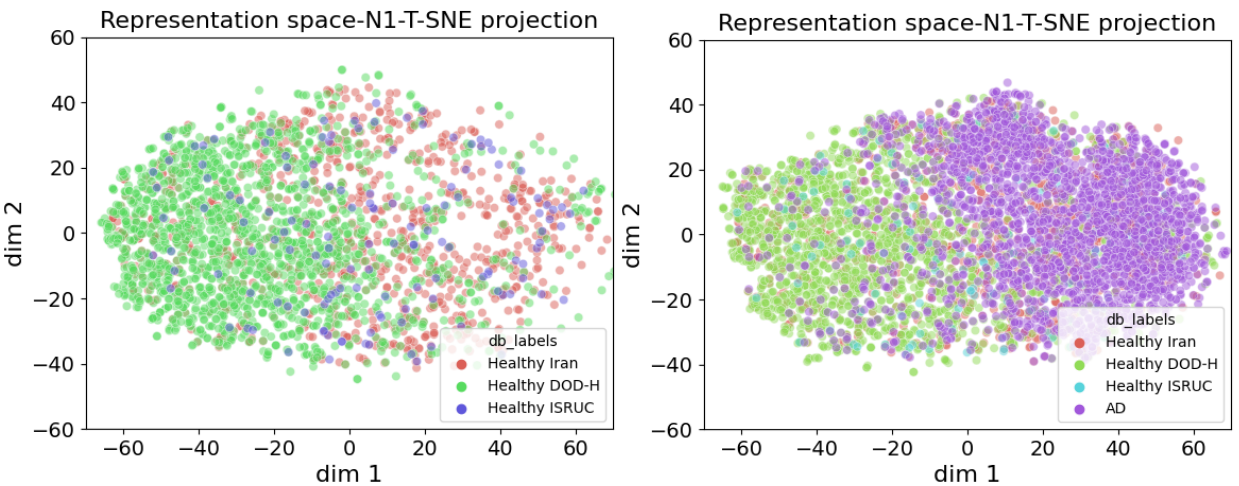}
        \caption{}
        \label{N1_tsne}
    \end{subfigure}

    \vspace{0.1cm}

    \begin{subfigure}{\textwidth}
        \centering
        \includegraphics[width=0.77\textwidth]{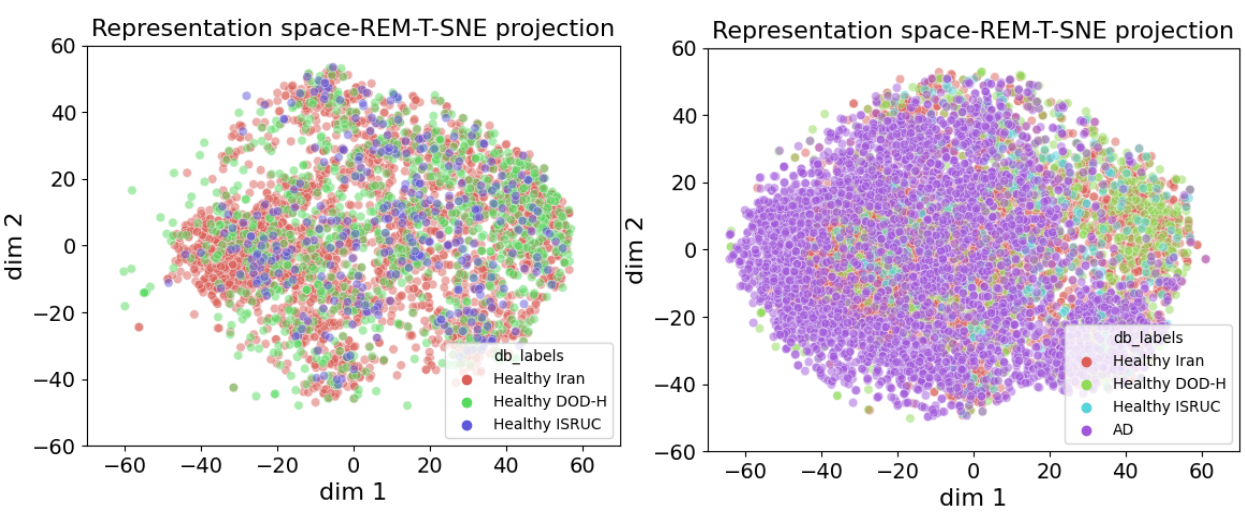}
        \caption{}
        \label{REM_tsne}
    \end{subfigure}

    \caption{T-SNE visualizations representing the post-training segregation of classes using features extracted by the SMATE model. The first column depicts the spatial distribution of healthy individuals, while the second column includes the entire dataset. Notably, there is a distinction between healthy and AD cases in both the N1 stage (a) and REM stage (b).}
    \label{tsne}
\end{figure*}

Fig.\ref{tsne} displays the t-SNE mappings of the N1 and REM sleep stages for a fully supervised SMATE model. The visualizations on the left in both Fig.~\ref{N1_tsne} and Fig.~\ref{REM_tsne} depict the post-training state, showing the distribution of individual samples from the combined healthy control (HC) databases. The visualizations on the right demonstrate the separation of the dataset into groups of healthy individuals and those diagnosed with Alzheimer's Disease (AD).

Post-training t-SNE visualizations, generated using features extracted by the SMATE model, reveal a distinct pattern where data points belonging to the same class (healthy vs. AD) tend to cluster together even coming from different databases, effectively distinguishing them from the opposing class. Thus, we conclude that we do not suffer from domain misalignment. These visual representations validate the effectiveness of SMATE in differentiating between classes, which aligns with the quantitative results previously discussed. 

However, some anomalies are observed within these projections. Certain data points are positioned outside their expected clusters, suggesting potential misclassification. These observations may be attributed to the inherent nature of t-SNE, which consolidates all dimensions into a 2-dimensional plane, potentially leading to the loss of certain aspects, such as depth.


\section{Discussion}
\label{sec: discussion}

This study makes a significant contribution to the field of Alzheimer's Disease (AD) detection through the analysis of sleep EEG signals using semi-supervised deep learning techniques. Our research has shown that the SMATE model, with its semi-supervised approach, achieves a considerable accuracy in detecting AD, particularly in scenarios characterized by a scarcity of labeled data. This is a promising development, as it suggests that semi-supervised learning can effectively leverage the limited labeled data typically available in clinical settings, supplemented by a larger volume of unlabeled data. 

The performance of SMATE, TapNet, XCM, and HMM was compared, demonstrating the efficacy of semi-supervised models in capturing complex signal patterns and variations indicative of AD. Notably, the SMATE model's ability to differentiate between healthy and AD cases in various sleep stages demonstrates the potential of sleep EEG signals as viable biomarkers for early AD detection. This is particularly crucial given the subtle and often pre-symptomatic nature of AD, where early detection can significantly impact the effectiveness of interventions. 

Our experiments with varying proportions of labeled data for SMATE and TapNet revealed several key insights, notably that the classification on the N1 stage consistently emerged as the most stable across all scenarios. SMATE, efficiently extracted spatial and temporal features from the sleep EEG data, confirming the importance of preserving channel order in the input signals. Its consistent performance across all sleep phases, with minimal labeled data increments, demonstrates its robustness and learning capability. This finding is visually supported by t-SNE projections, which illustrated effective class segregation and the seamless integration of various healthy individual databases without compromising classification accuracy. Thus, this validates our preprocessing strategy and the models' ability to discern disease-specific patterns. 


Moreover, the results of the ablation studies highlight the critical importance of both spatial and temporal feature extraction. These components are essential for capturing knowledge of semi-supervised patterns, contributing significantly to contextual understanding and the interpretation of temporal sequences. The removal of either block significantly impacts performance, with spatial extraction particularly affecting the overall effectiveness of both SMATE and TapNet. Conversely, XCM, which benefits from robust, fully supervised learning, appears to be less dependent on these layers, consistently demonstrating superior performance across different stages. This difference underlines the need for careful consideration of model architecture and the importance of features in designing effective tools for AD detection. It also points to the potential of combining different model types or features to optimize performance. 

While XCM and SMATE demonstrated impressive metrics, certain factors could affect these outcomes. The inclination of the AD database towards older age groups suggests that the models might be recognizing age-related signal patterns across various sleep phases rather than AD-specific characteristics. Considering this, we conducted age-based patient classification tests, segmenting the signals into three distinct age categories and compiling a balanced database of healthy and AD patients. Contrary to expectations, the models were unable to achieve satisfactory classification, often confusing the signals across different age groups. This outcome indicates the models' capability to discern Alzheimer's-related patterns, effectively distinguishing them from age-related variations within the combined databases. 

Furthermore, experiments were conducted on the complete signal without considering division by sleep phases. The outcomes revealed inferior metrics, underscoring the significance of independent stage analyses.

\section{Conclusion}
\label{sec: conclusion}

In conclusion, this research represents an important step toward more accurate and efficient early detection of Alzheimer's Disease. By leveraging advanced machine learning techniques and the rich information present in sleep EEG signals, it is hoped that this work will lead to better diagnostic tools and, ultimately, more effective treatments for AD. The integration of semi-supervised learning models like SMATE in clinical practice could significantly enhance our ability to detect AD early, offering a critical window for intervention and potentially altering the disease's trajectory for many individuals. 

Our detailed examination of each sleep phase independently optimized prediction accuracy. Visual tools like t-SNE projections were instrumental in elucidating the model's functionality and pinpointing areas of potential misclassification. Furthermore, this study has underscored the critical importance of integrating spatial and temporal features for successful signal analysis. 

Future research should focus on utilizing consistent and more diverse databases, refining model architectures, validating models against new datasets, and automating sleep phase classification. Additionally, a comprehensive examination of signal artifacts' impact on model performance may provide further insights. Advancements in these areas could lead to more precise and efficient tools for AD detection, substantially contributing to patient care and disease management.


\section*{Acknowledgements}

The work was partially supported by the Ministerio de Ciencia, Innovación y Universidades, Agencia Estatal de Investigación, under grants PID2019-109820RB-I00 and PID2021-123182OB-I00, MCIN/AEI/10.13039/501100011033, cofinanced by European Regional Development Fund (ERDF), "A way of making Europe." Pablo M. Olmos also acknowledges the support by the Comunidad de Madrid under grants IND2022/TIC-23550 and ELLIS Unit Madrid.

The authors wish to extend their gratitude to Manuel Sanchez de la Torre for the fruitful discussions (IRBLleida-Hospital Universitari Arnau de Vilanova and Santa Maria, Lleida) and Montse Pujol, Rafaela Vaca, Olga Minguez, (IRBLleida-Hosptial Universitari Arnau de Vilanova and Santa Maria, Lleida) for the PSG database manual scoring. 

The authors utilized the language model ChatGPT developed by OpenAI for assistance in the drafting of this paper.

\section*{References}
\bibliographystyle{IEEEtran}
\bibliography{references}

\begin{thebibliography}{10}
\providecommand{\url}[1]{#1}
\csname url@samestyle\endcsname
\providecommand{\newblock}{\relax}
\providecommand{\bibinfo}[2]{#2}
\providecommand{\BIBentrySTDinterwordspacing}{\spaceskip=0pt\relax}
\providecommand{\BIBentryALTinterwordstretchfactor}{4}
\providecommand{\BIBentryALTinterwordspacing}{\spaceskip=\fontdimen2\font plus
\BIBentryALTinterwordstretchfactor\fontdimen3\font minus \fontdimen4\font\relax}
\providecommand{\BIBforeignlanguage}[2]{{%
\expandafter\ifx\csname l@#1\endcsname\relax
\typeout{** WARNING: IEEEtran.bst: No hyphenation pattern has been}%
\typeout{** loaded for the language `#1'. Using the pattern for}%
\typeout{** the default language instead.}%
\else
\language=\csname l@#1\endcsname
\fi
#2}}
\providecommand{\BIBdecl}{\relax}
\BIBdecl

\bibitem{WorldHealthOrganization}
``Dementia. {W}orld {H}ealth {O}rganization ({WHO}).'' Mar 2023, \url{https://www.who.int/news-room/fact-sheets/detail/dementia}.

\bibitem{DONOSO2003}
A.~Donoso, ``\BIBforeignlanguage{es}{La enfermedad de {A}lzheimer},'' \emph{\BIBforeignlanguage{es}{Revista {C}hilena de {N}euro-psiquiatria}}, vol.~41, pp. 13 -- 22, 11 2003.

\bibitem{proteinas_ad}
M.~H. Janeiro, C.~G. Ardanaz, N.~Sola-Sevilla, J.~Dong, M.~Cortés-Erice, M.~Solas, E.~Puerta, and M.~J. Ramírez, ``Biomarcadores en la enfermedad de {A}lzheimer,'' \emph{Advances in Laboratory Medicine / Avances en Medicina de Laboratorio}, vol.~2, no.~1, p. 39–50, 2021.

\bibitem{ref9}
A.~G. Andrade, O.~M. Bubu, A.~W. Varga, and R.~S. Osorio, ``The relationship between obstructive sleep apnea and {A}lzheimer’s disease,'' \emph{Journal of Alzheimer’s Disease}, vol.~64, no.~s1, pp. S255--S270, 2018.

\bibitem{gaeta2020prevalence}
A.~M. Gaeta, I.~D. Ben{\'\i}tez, C.~Jorge, G.~Torres, F.~Dakterzada, O.~Minguez, R.~Huerto, M.~Pujol, A.~Carnes, M.~Dalmases \emph{et~al.}, ``Prevalence of obstructive sleep apnea in alzheimer’s disease patients,'' \emph{Journal of neurology}, vol. 267, pp. 1012--1022, 2020.

\bibitem{azami2023eeg}
H.~Azami, S.~Moguilner, H.~Penagos, R.~A. Sarkis, S.~E. Arnold, S.~N. Gomperts, and A.~D. Lam, ``Eeg entropy in rem sleep as a physiologic biomarker in early clinical stages of alzheimer’s disease,'' \emph{Journal of Alzheimer's Disease}, no. Preprint, pp. 1--16, 2023.

\bibitem{7488219}
S.~Afshari and M.~Jalili, ``Directed functional networks in alzheimer's disease: Disruption of global and local connectivity measures,'' \emph{IEEE Journal of Biomedical and Health Informatics}, vol.~21, no.~4, pp. 949--955, 2017.

\bibitem{6845193}
K.~A.~I. Aboalayon, H.~T. Ocbagabir, and M.~Faezipour, ``{E}fficient sleep stage classification based on {EEG} signals,'' in \emph{{IEEE} {L}ong {I}sland {S}ystems, {A}pplications and {T}echnology ({LISAT}) {C}onference 2014}, 2014, pp. 1--6.

\bibitem{9440810}
M.~Tanveer, A.~H. Rashid, M.~A. Ganaie, M.~Reza, I.~Razzak, and K.-L. Hua, ``Classification of {A}lzheimer’s {D}isease using ensemble of {D}eep {N}eural {N}etworks trained through {T}ransfer {L}earning,'' \emph{IEEE Journal of Biomedical and Health Informatics}, vol.~26, no.~4, pp. 1453--1463, 2022.

\bibitem{10271565}
D.~Klepl, F.~He, M.~Wu, D.~J. Blackburn, and P.~Sarrigiannis, ``Adaptive {G}ated {G}raph {C}onvolutional {N}etwork for {E}xplainable {D}iagnosis of {A}lzheimer’s {D}isease {U}sing {EEG} {D}ata,'' \emph{IEEE Transactions on Neural Systems and Rehabilitation Engineering}, vol.~31, pp. 3978--3987, 2023.

\bibitem{TAUTAN2021102081}
A.-M. Tăuţan, B.~Ionescu, and E.~Santarnecchi, ``Artificial intelligence in neurodegenerative diseases: {A} review of available tools with a focus on machine learning techniques,'' \emph{Artificial Intelligence in Medicine}, vol. 117, p. 102081, 2021.

\bibitem{DAtri_Scarpelli_Gorgoni_Truglia_Lauri_Cordone_Ferrara_Marra_Rossini_De_Gennaro_2021}
A.~D’Atri, S.~Scarpelli, M.~Gorgoni, I.~Truglia, G.~Lauri, S.~Cordone, M.~Ferrara, C.~Marra, P.~M. Rossini, and L.~De~Gennaro, ``{EEG} alterations during wake and sleep in mild cognitive impairment and alzheimer’s disease,'' \emph{iScience}, vol.~24, no.~4, p. 102386, 2021.

\bibitem{Geng_Wang_Fu_Zhang_Yang_An_2022}
D.~Geng, C.~Wang, Z.~Fu, Y.~Zhang, K.~Yang, and H.~An, ``Sleep {EEG}-based approach to detect {M}ild {C}ognitive {I}mpairment,'' \emph{Frontiers in Aging Neuroscience}, vol.~14, 2022.

\bibitem{Azami_Moguilner_Penagos_Sarkis_Arnold_Gomperts_Lam_2023}
H.~Azami, S.~Moguilner, H.~Penagos, R.~A. Sarkis, S.~E. Arnold, S.~N. Gomperts, and A.~D. Lam, ``{EEG} entropy in {REM} sleep as a physiologic biomarker in early clinical stages of {A}lzheimer’s {D}isease,'' \emph{Journal of Alzheimer’s Disease}, vol.~91, no.~4, p. 1557–1572, 2023.

\bibitem{ASSM}
B.~Duce, C.~Rego, J.~Milosavljevic, and C.~Hukins, ``The {AASM} recommended and acceptable {EEG} montages are comparable for the staging of sleep and scoring of {EEG} {A}rousals,'' \emph{Journal of Clinical Sleep Medicine}, vol.~10, no.~07, p. 803–809, 2014.

\bibitem{27}
R.~Wang, J.~Wang, H.~Yu, X.~Wei, C.~Yang, and B.~Deng, ``Power spectral density and coherence analysis of alzheimer’s {EEG},'' \emph{Cognitive Neurodynamics}, vol.~9, no.~3, p. 291–304, 2014.

\bibitem{30}
S.~Khalighi, T.~Sousa, J.~M. Santos, and U.~Nunes, ``{ISRUC}-sleep: {A }comprehensive public dataset for sleep researchers,'' \emph{Computer Methods and Programs in Biomedicine}, vol. 124, p. 180–192, 2016.

\bibitem{dodh}
A.~Guillot, F.~Sauvet, E.~During, and V.~Thorey, ``Dreem {O}pen {D}atasets: {M}ulti-{S}cored {S}leep {D}atasets to {C}ompare {H}uman and {A}utomated {S}leep {S}taging,'' \emph{IEEE Transactions on Neural Systems and Rehabilitation Engineering}, vol.~PP, pp. 1--1, 2020.

\bibitem{32}
M.~Rezaei, H.~Mohammadi, and H.~Khazaie, ``{EEG/EOG/EMG} data from a cross sectional study on psychophysiological insomnia and normal sleep subjects,'' \emph{Data in Brief}, vol.~15, p. 314–319, 2017.

\bibitem{Terry_Anderson_Horne_2004}
J.~R. Terry, C.~Anderson, and J.~Horne, ``Nonlinear analysis of {EEG} during {NREM} sleep reveals changes in functional connectivity due to natural aging,'' \emph{Human Brain Mapping}, vol.~23, no.~2, pp. 73--84, Jun. 2004.

\bibitem{S-to-N}
H.~Huang, J.~Zhang, L.~Zhu, J.~Tang, G.~Lin, W.~Kong, X.~Lei, and L.~Zhu, ``{EEG}-based sleep staging analysis with functional connectivity,'' \emph{Sensors}, vol.~21, no.~6, p. 1988, 2021.

\bibitem{waves_freq}
A.~Coatanhay, L.~Soufflet, L.~Staner, and P.~Boeijinga, ``{EEG} source identification: {F}requency analysis during sleep,'' \emph{Comptes Rendus Biologies}, vol. 325, no.~4, p. 273–282, 2002.

\bibitem{smate}
J.~Zuo, K.~Zeitouni, and Y.~Taher, ``Smate: {S}emi-supervised spatio-temporal representation learning on multivariate time series,'' in \emph{2021 IEEE International Conference on Data Mining (ICDM)}.\hskip 1em plus 0.5em minus 0.4em\relax IEEE, 2021, pp. 1565--1570.

\bibitem{tapnet}
X.~Zhang, Y.~Gao, J.~Lin, and C.-T. Lu, ``Tap{N}et: {M}ultivariate time series classification with attentional prototypical network,'' \emph{Proceedings of the AAAI Conference on Artificial Intelligence}, vol.~34, no.~04, p. 6845–6852, 2020.

\bibitem{xcm}
K.~Fauvel, T.~Lin, V.~Masson, E.~Fromont, and A.~Termier, ``{XCM}: {A}n explainable convolutional neural network for {M}ultivariate time series classification,'' \emph{Mathematics}, vol.~9, no.~23, p. 3137, 2021.

\bibitem{git}
F.~Moreno-Pino, E.~Sükei, P.~M. Olmos, and A.~Artés-Rodríguez, ``{P}y{HHMM}: A {P}ython {L}ibrary for {H}eterogeneous {H}idden {M}arkov {M}odels,'' 2022.

\bibitem{anova}
H.-Y. Kim, ``Analysis of variance ({ANOVA}) comparing means of more than two groups,'' \emph{Restorative dentistry \& endodontics}, vol.~39, no.~1, pp. 74--77, 2014.

\bibitem{tsne2}
A.~A. Awan, ``Introduction to {T-SNE}: {N}onlinear dimensionality reduction and {D}ata {V}isualization,'' Mar 2023, \url{https://www.datacamp.com/tutorial/introduction-t-sne}.

\end{thebibliography}


\appendices
\section{}

  \label{FirstAppendix}


\begin{table}[H]
        \scalebox{1.2}{
            \begin{tabular}{| p{3.5cm} | p{3cm} | } \hline
                \textbf{Hyperparameters} & \textbf{Values}\\ [0.5ex] \hline
                \textbf{Signal length} &  1280\\
                \textbf{Nº of epochs} & 1000\\
                \textbf{Loss} & Categorical cross-entropy\\
                \textbf{Neighboring window sizes } &  8, 5, 3\\
                \textbf{Hidden SMBs dimension} &  2\\
                \textbf{Conv1D output dimension of the encoder} &  128, 256, 128\\
                \textbf{Pooling size} &  128\\
                \textbf{Output GRUs dimensions} &  128\\
                \textbf{FC layers output dimension of the encoder} & 128   \\ \hline
            \end{tabular}
        }
        \caption{Parameter selection used for the SMATE model.}
        \label{subtab:smate_parameters}
\end{table}

\begin{table}[H]
        \scalebox{1.2}{
            \begin{tabular}{| p{3.5cm} | p{3cm} | } \hline
                \textbf{Hyperparameters} & \textbf{Values}\\ [0.5ex]\hline
                \textbf{Signals length} &  1280\\
                \textbf{Number of epochs} & 3000\\
                \textbf{Loss} & Categorical cross-entropy\\
                \textbf{Parameters for random projection configuration} &  -1, 3 \\
                \textbf{CNN filters} &  256, 256, 128\\
                \textbf{CNN kernels} &  8, 5, 3\\
                \textbf{Layer settings of mapping function} &  500, 300 \\
                \textbf{Weight decay (L2 loss on parameters)} & $10^{-3}$\\ \hline
            \end{tabular}
        }
        \caption{Parameter selection used for the TapNet model.}
        \label{subtab:tapnet_parameters}
\end{table}

\begin{table}[H]
        \scalebox{1.2}{
            \begin{tabular}{| p{3.5cm} | p{3cm} | } \hline
                \textbf{Hyperparameters} & \textbf{Values}\\ [0.5ex] \hline
                \textbf{Signals length} &  1280\\
                \textbf{Number of epochs} & 10\\
                \textbf{Window sizes (\% of MTS used to extract features)} &  0.3\\
                \textbf{Loss} & Categorical cross-entropy \\
                \textbf{Conv2D and Conv1D padding} &  Same\\
                \textbf{Conv2D and Conv1D filters} &  128\\
                \textbf{Activation} &  ReLu\\
                \textbf{Output activation layer} & Softmax   \\ \hline
            \end{tabular}
        }
        \caption{Parameter selection used for the XCM model.}
        \label{subtab:xcm_parameters}
\end{table}

\begin{table}[H]
        \scalebox{1.2}{
            \begin{tabular}{| p{3.5cm} | p{3cm} | } \hline
                \textbf{Hyperparameters} & \textbf{Values}\\ [0.5ex] \hline
                \textbf{Signals length} &  1280\\
                \textbf{Features extracted} &  Mean, std, kurtosis, Hjorth parameters\\
                \textbf{Number of hidden states} &  5\\
                \textbf{Number of emissions} &  4\\
                \textbf{Type of HMM emissions} &  Gaussian\\
                \textbf{Covariance type} &  Diagonal\\
                \textbf{Number of iterations} & 1000   \\ 
                \textbf{Convergence threshold} &  0.001\\
                \textbf{Convergence iterations} &  5\\\hline
            \end{tabular}
        }
        \caption{Parameter selection used for the HMM model.}
        \label{subtab:hmm_parameters}
\end{table}

\begin{table}[H]
    \centering    
    \scalebox{1.2}{ 
        \begin{tabular}{| c | c | c |} \hline
            \textbf{Model} & Without artifacts & With artifacts   \\ \hline
            \multicolumn{3}{|c|}{\textbf{Accuracy}}\\ \hline
            \textbf{SMATE} & \textbf{0.809 ± 0.041}  & \textbf{0.849 ± 0.057}  \\
            \textbf{TapNet} & 0.768 ± 0.056 & 0.765 ± 0.086 \\ \hline
            \multicolumn{3}{|c|}{\textbf{ROC/AUC}}\\ \hline
            \textbf{SMATE} & \textbf{0.809 ± 0.041}  & \textbf{0.849 ± 0.057}  \\
            \textbf{TapNet} & 0.768 ± 0.056 & 0.765 ± 0.086 \\ \hline
        \end{tabular}}
            \caption{Comparison of test accuracy and ROC/AUC with their standard deviation (averaged across 5 folds) for semi-supervised models applied to the PSG dataset without segmentation into sleep stages. The analysis explores the importance of this division and evaluates the impact of artifacts in the final metrics.}
    \label{tab:all_signals_results}
\end{table}

\begin{table*}[htb]
    \centering    
    \scalebox{1.2}{ 
        \begin{tabular}{| c | c | c | c | c |} \hline
            \textbf{Model} & N1 & N2  & N3 & REM  \\ \hline
            \multicolumn{5}{|c|}{\textbf{Accuracy}}\\ \hline
            \textbf{SMATE} & \textbf{0.354 ± 0.224}  & \textbf{0.387 ± 0.205} & \textbf{0.511 ± 0.214} & 0.403 ± 0.105 \\
            \textbf{TapNet} & 0.295 ± 0.135  & 0.369 ± 0.120 & 0.497 ± 0.144 & \textbf{0.544 ± 0.081} \\ \hline
            \multicolumn{5}{|c|}{\textbf{F1-score}}\\ \hline
            \textbf{SMATE} & \textbf{0.336 ± 0.215}  & \textbf{0.394 ± 0.230} & 0.516 ± 0.197 & 0.349 ± 0.131  \\
            \textbf{TapNet} & 0.312 ± 0.109  & 0.372 ± 0.137 & \textbf{0.521 ± 0.195} & \textbf{0.507 ± 0.105} \\ \hline
        \end{tabular}}
            \caption{Comparison of test accuracy and F1-score, along with their standard deviation (averaged across 5 folds), for semi-supervised models applied to the classification of PSG signals across three age groups. The analysis emphasizes the models' ability to distinguish AD-related patterns rather than age-related patterns.}
    \label{tab:ages_results}
\end{table*}

\section{In-depth SMATE}
    \label{Deep dive into SMATE}

As mentioned in the main text, SMATE model employs an autoencoder framework that translates MTS samples from their native space into a condensed embedding space, and a specialized regularization approach for the model optimization. With respect the key components, their structure include:

\paragraph{Spatial Modeling Block (SMB)}

Analysis of the spatial interactivity among channels. It inputs a time series (h), characterized by its T (length) and d (number of channels). As depicted in Fig.~\ref{smb}, the block undergoes sequential 1D average pooling operations on each variable. This encodes neighboring sample information into $(S_H)$ which is then processed by two fully connected (FC) layers, facilitating the vertical interactions. This derivation offers spatial interaction weights for every one-dimensional segment, mirroring the dimensions of the input. The sequence culminates with an element-wise multiplication for each initial time series segment~\cite{smate}.

\paragraph{Spatio-Temporal Encoding on MTS}

Spatial extraction involves a trio of SMB blocks, each succeeded by a convolution 1D (Conv1D) and a batch normalization (BN) layer empowered by ReLu activation. The outcome is a representation $h (S)  \in  \mathbb{R}  ^{L  x  d_c}$, being  $d_c$ the filter number, and $L = T/P$, being $P$ the pooling size. 

Conversely, the encoder temporal extraction block incorporates three GRUs. The output of the three blocks, followed by a final 1D pooling layer, yields a representation $h (T) \in \mathbb{R} ^{L x d_g}$, where $d_g$ indicates the GRUs output dimension.

Both the temporal and spatial block outcomes are concatenated and subsequently processed through two FC layers, deriving the spatio-temporal embedding.

\paragraph{Joint Model Optimization}

The autoencoder framework exhibits a sparse distribution focus on the restoration performance of the embeddings. To address this, SMATE introduces a three-step joint model optimization that integrates with temporal reconstructions, combining both labeled and unlabeled samples. The temporal reconstruction loss is defined as: 
\begin{equation}
    L_{R} =  \sum_t || x_{t} - \tilde{x} ||
\end{equation}
where $x$ and $\tilde{x}$ corresponds to the raw and reconstructed samples. Consequently, the three step regularization approaches the class-specific clusters for each embedding space to the virtual centroids during the training phase:

\begin{itemize}
    \item \textbf{Step 1 (Supervised Centroids Initialization:)} Class centroids are initialized as the average of all representations within the class-specific embedding space of the labeled training set. 

    \item \textbf{Step 2 (Supervised Centroids Adjustment:)} The probabilities derived from the samples distances to each class, learned through supervised learning, serve to refine the centroids, as they assess the contribution of each of these samples on the centroid’s decision. The metric utilized in this phase is the inverse of the Euclidean distance (ED). 

    \item \textbf{Step 3 (Unsupervised Centroids Adjustment:)} The labels are propagated via distance-based class probability to the unlabeled samples, so this samples are assimilated into the class-specific cluster with the highest probability, being able to adjust the centroids. 
\end{itemize}

 The regularization loss is subsequently computed as:

\begin{equation}
    L_{Reg}(\theta) = - \sum_k \log W_{\theta}(y = k|x)
\end{equation}

Here, $W_{\theta}(y = k|x)$ denotes the weights associated with the labeled samples or the distance-based probabilities of the unlabeled samples, being x the raw samples, and k the samples class. 
The comprehensive optimization can then be represented as:
\begin{equation}
    \min_\theta(L_R + \lambda L_{Reg})
\end{equation}
In this equation, $\lambda$ serves as a hyperparameter $\geq$ 0.

As a concluding step, a linear kernel SVM is trained using the embeddings derived from the preceding blocks, culminating in a finalized classification.

\end{document}